\documentclass[]{interact}

\usepackage{epstopdf}
\usepackage[caption=false]{subfig}

\usepackage{hyperref}
\usepackage{physics}
\usepackage{placeins}

\usepackage[numbers,sort&compress]{natbib}
\bibpunct[, ]{[}{]}{,}{n}{,}{,}
\renewcommand\bibfont{\fontsize{10}{12}\selectfont}
\makeatletter
\def\NAT@def@citea{\def\@citea{\NAT@separator}}
\makeatother

\theoremstyle{plain}

\theoremstyle{definition}

\theoremstyle{remark}

\begin{document}

\articletype{ARTICLE TEMPLATE}

\title{Inderdigitation, double twist, and topological defects in a system of hard lollipops}

\author{
\name{P. Kubala\textsuperscript{a}\thanks{CONTACT P. Kubala. Email: piotr.kubala@doctoral.uj.edu.pl} and Micha\l{} Cie\'sla\textsuperscript{a}}
\affil{\textsuperscript{a}Marian Smoluchowski Institute of Theoretical Physics, Jagiellonian University, Krak\'ow}
}

\maketitle

\begin{abstract}
Using hard particle Monte Carlo simulations, we studied the three-dimensional system consisting of identical, lollipop-like particles. Each lollipop was built of five identical, tangent balls placed along a line and one larger ball at one side of the particle and modeled the RM734 molecule, for which ferroelectric and splay nematics were recently discovered in the experiment. Although our model did not recreate these phases, we observed inherently polar type A and C interdigitated smectics. Moreover, an intriguing, isotropic phase consisting of double-twisted clusters joined by planar defects was formed for a moderate packing fraction and ball diameters ratio.
\end{abstract}


\section{Introduction}

Although there is no fundamental reason against the formation of nematic ferroelectric phase in liquids made of elongated molecules with large dipole moments \cite{Palffy1988, Rosseto2020}, it has not been observed until recently \cite{Mertelj2013, Shuai2016, Sebastian2020, Chen2020}. Therefore, the discovery started a discussion of the molecular origin of such an alignment of molecules \cite{Mandle2021, Mandle2022}. The most straightforward approach requires the violation of the up-down symmetry of a single molecule to provoke such symmetry breaking in the whole phase. In this context, two main options are considered. Symmetry can be violated by either the electrostatic charge distribution along a molecule or its shape. Here, we focus on the second option. It is worth noting that the packing entropy of anisotropic molecules without any additional soft interactions allowed explaining the existence of a variety of liquid crystalline phases starting from the most basic ones like nematics and smectics \cite{Allen2017} and ending on twist-bent phases discovered in the second decade of 21\textsuperscript{st} century \cite{Greco2015, Chiappini2021, Kubala2022}.
On the other hand, studies of crystalline phases composed of hard pear- and taper-like molecules show that the splay order, even if observed, cannot propagate uniformly in the whole system \cite{Schonhofer2017, Schonhofer2020, Kubala2023}. To optimize the packing fraction, domains of opposite polarities and splay vectors are formed, introducing defects. Point defects in liquid crystalline phases are typically a consequence of some impurities or special preparation of the liquid crystal sample, as they significantly raise the local free energy. On the other hand, their stability can be assured by the topological charge they carry \cite{deGennesBook}. Linear defects are observed in blue phases, where the optimal placing on neighboring molecules requires a non-zero twist between them \cite{Wright1989, Kikuchi2002}. For some specific value of this twist, molecules form so-called double-twisted tubes where their orientations deviate from the tube axis the more, the farther from it the molecule is. These tubes can form some regular patterns (BPI and BPII) or an isotropic structure (BPIII); however, in all these cases, we observe defects at the border of two neighboring tubes, as the mean local orientation is undefined there. Finally, as mentioned above, planar defects are expected in polar splay phases composed of hard taper-like molecules.

In this study, we want to explore the phase diagram of the system built of lollipop-shaped particles using Monte Carlo simulations. In such a system, one can expect the formation of polar phases, which have been observed for similar taper-like molecules but only in the crystalline phase \cite{Kubala2023}. It is well known that even a slight change in a hard particle's shape can significantly impact the self-assembly structures. One of the best examples are systems composed of hard ellipsoids and similar hard spherocylinders, where a smectic A phase is observed only for the second particle type \cite{Frenkel1984, Veerman1990}. It is worth noticing that a similar effect is also observed in the experimental studies of polar nematics when seemingly insignificant changes are introduced in the molecular structure. For example, although the ferroelectric nematic was observed for RM734 molecules, the phase disappears when the NO\textsubscript{2} group at the end of the molecule is replaced by the CN group \cite{Mandle2017a, Mandle2021}. In the same spirit, we investigate here a shape that is similar to previously studied taper-like particles \cite{Gregorio2016, Kubala2023, Kubala2023sm} to find out how it affects the existence of particular phases and their range of stability compared to the former systems. The lollipop model is known in the literature, and it was previously studied in two dimensions, where it exhibits standard nematic and various interdigitated smectic A phases, depending on the fine details of the particles' shape \cite{Perez2017}.

\section{Model and methods}

\subsection{Model particle}

\begin{figure}[hb]
    \centering
    \includegraphics[width=0.4\linewidth]{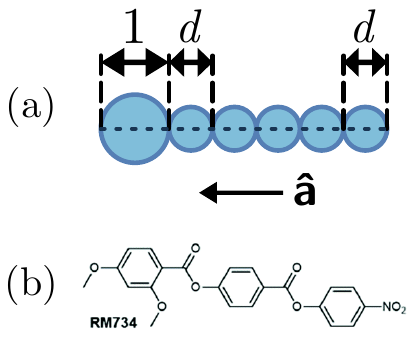}
    \caption{(a) Lollipop particle model: six colinear tangent spherical beads; the first one has diameter 1, while the five remaining ones have a diameter $d \in [0.4, 1.0]$, which is a parameter. Here, $\vu{a}$ is the long molecular axis and the polarization axis. (b) Taper-shaped RM734 molecule, for which ferroelectic and splay nematics were observed experimentally. Reprinted from Ref.~\cite{Mandle2022} with the permission of the Royal Society of Chemistry.}
    \label{fig:lollipop}
\end{figure}

To recreate the shape of the RM734 molecule, we use a hard-core toy model created from six co-linear tangent beads. Among them, five are identical and have a diameter of $d$, while the last one at the endpoint has a diameter of 1. We restrict d to the range $d \in [0.4, 1]$. Consequently, the shape resembles a lollipop. Its length is $1 + 5d$, while the average ball diameter is $(1 + 5d)/6$, thus, the aspect ratio is constant and equal to 6. Alternatively, if one defines the shape's width as the diameter of the largest ball, the aspect ratio is equal to the length $1 + 5d$.

\subsection{Monte Carlo simulations}

The system was simulated using Monte Carlo method in the isothermal-isobaric $NpT$ ensemble \cite{Wood1968, Allen2017, Allen2019}. The initial states were prepared as follows. The system was first arranged in an antiferroelectric hexagonal closed-packed (hcp) configuration with layers along the $z$ axis. More precisely, the particles in each honeycomb layer had identical polarizations (the orientations of $\vu{a}$), while at the same time, the polarizations were opposite $\vu{a} \leftrightarrow -\vu{a}$ in the neighboring layers. Then, the layers were compressed by moving them maximally close to one another without introducing overlaps. As in the hcp configuration the neighboring layers are shifted, for $d \le 0.65$ interdigitated configurations were obtained, where the chains of five smaller balls of ``up'' and ``down'' particles were able to fit between themselves. On the contrary, for $d \ge 0.7$ the interdigitation in the hcp configuration was not possible. As for $d \le 0.65$ we use twelve initial layers, while for $d \ge 0.7$ -- six, each configuration had finally six layers (or bi-layers) after the compression. All systems consisted of $N = 4500-4800$ particles.

The configurations were then equilibrated for $5 \times 10^8$ full MC cycles under a relatively high pressure $p^* = 20$ corresponding to packing fraction $\eta = 0.5-0.55$. Here, $p^* = p V_P/(k_B T)$ is the reduced pressure rescaled by the temperature, which is a scale-independent parameter suitable for hard-core interactions where $p/T$ is the only independent thermodynamic variable. $V_P$ is the particle's volume and $k_B$ is the Boltzmann constant. The equilibrated configurations were then expanded to all target pressures $p^* \in [4, 16]$, covering the phase sequence down to the isotropic liquid. Initial high-pressure $p^* = 20$ configurations were discarded due to the high probability of being metastable jammed states. Additionally, for $d \ge 0.85$, the systems with $\eta = 0.45$ were compressed back up to the crystalline phase. Equilibration was extremely slow, requiring up to $10^9$ full MC cycles for the densest configurations under investigation. Due to that, we could not obtain the equilibrium crystalline phases in the range $d \in [0.4, 0.8]$ within the reasonable computational timeframe.

Each full MC cycle consisted of one volume move and $1.1N$ particle moves; particle move type was chosen randomly: rototranslation move with the probability $10/11$ and flip move with the probability $1/11$. In the volume move, the box height (along the $z$ axis) and the box base were scaled by two independent, logarithmically sampled factors. For the densest system, the most general triclinic box was used, and volume moves were performed by perturbing all box vectors by small, random displacements. Overlapping configurations were immediately rejected, while the remaining ones were accepted according to the Metropolis-Wood criterion \cite{Wood1968, Wood1968jcp}
\begin{equation}
    P = \min\qty{1, \exp(N \log \frac{V_1}{V_0} - \frac{p \Delta V}{k_B T})},
\end{equation}
where $V_0$ is the initial system volume, $V_1$ is the volume after the move and $\Delta V = V_1 - V_0$. A rototranslation move entailed displacing the particle by a random vector and rotating it around a random axis by a random, uniformly sampled clockwise or anticlockwise angle. Finally, during the flip move, the particle was rotated by $180^\circ$ degrees around the middle of its length so that the molecular axis changed sign $\vu{a} \leftrightarrow -\vu{a}$. All particle moves introducing overlaps were rejected, while the rest was accepted. The extends of volume and rototranslation moves were dynamically adjusted so that the acceptance ratio was around 15\%, which is the optimal value for hard-core systems \cite{Mbamala2002}.

\subsection{Order parameters and correlation functions} \label{sec:params}

We introduced a set of order parameters to facilitate the recognition of phases in the system. Nematic order can be measured by the nematic order parameter $\expval{P_2}$. It is obtained by first computing the snapshot-averaged $\vb{Q}$-tensor \cite{Vieillard1974, Eppenga1984}
\begin{equation}
    \vb{Q} = \frac{1}{N} \sum_{i=1}^{N} \frac{3}{2}\qty(\vu{a}_i \otimes \vu{a}_i - \frac{1}{3} \vb{I}),
\end{equation}
where $\cdots \otimes \cdots$ is the outer product and $\vb{I}$ is the identity matrix. Its eigenvalue with the highest magnitude is an instantaneous $P_2$ value, and the corresponding eigenvector is the system director. The $P_2$ parameter is then time-averaged (ensemble-averaged) over independent system snapshots. In the ideal nematic $\expval{P_2} = 1$, for isotropic liquid $\expval{P_2} = 0$, while the minimal value $\expval{P_2} = -0.5$ is reached if all particles are oriented randomly, but orthogonal to the director.

The presence of density undulation can be probed using the smectic order parameter \cite{deGennesBook}
\begin{equation}
    \expval{\tau} = \frac{1}{N} \expval{
        \max_{h,k,l}
        \abs{
            \sum_{i=1}^{N}
            \exp(\iota \vb{k}_{hkl} \vdot \vb{r}_i)
        }
    }.
\end{equation}
Here, $\iota$ is the imaginary unit, $\vb{r}_i$ is the middle of the $i$th particle's length, while $\vb{k}_{hkl}$ is smectic wavevector. The modulus $\abs{\cdots}$ is computed before averaging in order to eliminate the translational Goldstone mode. Averaging over the independent system snapshots is denoted as $\expval{\cdots}$ Allowed, compatible with periodic boundary conditions (PBC) values of $\vb{k}_{hkl}$ can be enumerated using the Miller indices $h$, $k$, $l$ \cite{Kittel2018}:
\begin{align}
    \vb{k}_{hkl} &= h \vb{g}_1 + k \vb{g}_2 + l \vb{g}_3, \\
    \vb{g}_i &= \frac{2\pi}{V} (\vb{b}_{i+1} \cp \vb{b}_{i+2}),
\end{align}
where $\vb{b}_1 \equiv \vb{b}_4$, $\vb{b}_2 \equiv \vb{b}_5$, and $\vb{b}_3$ are real-space box vectors and $\vb{g}_i$ are reciprocal box vectors. As the initial configurations had six layers, we scanned $h,k,l$ in the range $[-6, 6]$. The smectic order parameter $\expval{\tau}$ ranges from 0 for a uniform density to 1 for ideal layers with co-planar $\vb{r}_i$.

The formation of antiferroelectric, bilayer, interdigitated arrangement can be registered quantitatively by incorporating the polarization into a smectic-like order parameter
\begin{equation}
    \expval{\tau_p} = \frac{1}{N} \expval{
        \abs{
            \sum_{i=1}^{N}
            (\vu{a}_i \vdot \vu{n})
            \exp(\iota \vb{k}_{hkl} \vdot \vb{r}_i)
        }
    }.
\end{equation}
Here, the exponential function is multiplied by a projection of a long molecular axis onto the director $\vu{n}$. By using the same wavevector $\vb{k}_{hkl}$ as in $\tau$ we probe if each smectic layer separates into two interdigitated sub-layers with different polarizations. It is important to note that choosing the length's middle instead of the mass center for the computation of both smectic and polarization order parameters eliminates the artificial separation of ``up'' and ``down'' particles due to the mass center being closer to one end of the shape. The polarization order parameter $\expval{\tau_p}$ is equal 0 in an apolar phase and reaches 1 for ideal, equidistant layers with an alternating polarization.

More information on the structure of the polarization field can be given by the $S_{110}$ orientational correlation function \cite{Stone1978}
\begin{equation}
    S_{110}(Z) = \expval{\vu{a}_i \vdot \vu{a}_j}_{\abs{(\vb{r}_j - \vb{r}_i) \vdot \vu{z}} \approx Z}.
\end{equation}
The averaging is done over all pairs of particles whose distance along the $z$ axis (perpendicular to the layering) is around $Z$. The minimal value $S_{110} = -1$ is achieved for antiparallel alignment, while the maximal one $S_{110} = 1$ -- for parallel.

The tendency to form at least local tetratic or hexatic structures can be measured using the local bond order parameter \cite{Nelson2012}. In two dimensions, it is defined as
\begin{equation}
    \expval{\psi_\alpha} = \frac{1}{N} \expval{
        \sum_{i=1}^{N} \frac{1}{\alpha} \abs{
            \sum_{j\in\alpha\text{NN}(i)} \exp(\alpha\iota\theta_{ij})
        }
    },
\end{equation}
where $\alpha$ is rank, $\alpha\text{NN}(i)$ is a list of $\alpha$ nearest neighbors of $i$th particle, and $\theta_{ij}$ is the angle between the vector joining $i$th and $j$th particle, and a constant, arbitrary direction on the plane. For $\alpha = 4$ the tetratic order is measured, while for $\alpha = 6$ -- the hexatic. One can generalize the bond order parameter to three-dimensional layered systems by projecting all particles onto their nearest layer and progressing with the original two-dimensional definition. It is important to note that, due to computing the modulus before averaging, local bond order parameters are not equal 0 in disordered systems, but rather $\expval{\psi_4} \approx 0.45$ and $\expval{\psi_6} \approx 0.37$. Ideal tetratic and hexatic configurations render maximal $\expval{\psi_\alpha} = 1$ values for their respective ranks $\alpha$.

It is important to note that usually, the bond order parameter is defined in a global fashion, where the averaging over the particles and the modulus are swapped:
\begin{equation}
    \expval{\Psi_\alpha} = \frac{1}{N} \expval{
        \abs{
            \sum_{i=1}^{N} \frac{1}{\alpha} 
            \sum_{j\in\alpha\text{NN}(i)} \exp(\alpha\iota\theta_{ij})
        }
    }.
\end{equation}
Then, the global alignment of bonds is also probed. However, such global order parameter $\Psi_\alpha$ is prone to artificial lowering due to the formation of domains, thus, we revert to using the local version $\psi_\alpha$.

As the results will show, we observe a globally isotropic phase but with local double-twisted clusters. To detect them, we devise a complex $\chi$ order parameter, which is sensitive to the non-zero curl of the director field, consequently called the \textit{curl order parameter}. For $i$th particle, we define it as
\begin{equation} \label{eq:chi}
    \chi_i^{(R)} = \expval{-\iota \exp(\iota \beta_{ij})}_{j: \norm{\vb{r}_{ij}} \le R},
\end{equation}
where $\beta_{ij}$ is the directed angle between the vector $\vb{r}_{ij} = \vb{r}_j - \vb{r}_i$ joining centers of $i$th and $j$th particle, and the sign-aligned axis $(\vu{a}_i \vdot \vu{a}_j / \abs{\vu{a}_i \vdot \vu{a}_j}) \vu{a}_j$ of the $j$th particle, both projected onto the plane perpendicular to $\vu{a}_i$. The averaging is performed over all neighbors distant from the $i$th particle not further than $R$. The maximal distance $R$ controls the coarse-graining -- we choose it to be comparable with the length of the particle; it is dropped in the latter part of the manuscript $\chi_i^{(R)} \equiv \chi_i$. The maximal magnitude $\abs{\chi_i} = 1$ is reached when the angle $\beta_{ij}$ remains constant, i.e., for a rotationally symmetric director field around the central particle. When $\beta_{ij} = \pm \pi/2$, $\chi_i$ is equal $\pm 1$ and the field lines are concentric circles. Local deformations of a double-twisted type can be probed by analyzing the ensemble-averaged histogram of $\psi$ on a complex plane. A single sharp maximum should be observed around $\chi = 0$ for a uniform system. On the other hand, in the presence of double-twisted clusters, two maxima in $\chi = \pm z$, $z = \abs{z}e^{\iota \phi}$ will appear.

\section{Results}

\begin{figure}[htb]
    \centering
    \includegraphics[width=0.6\linewidth]{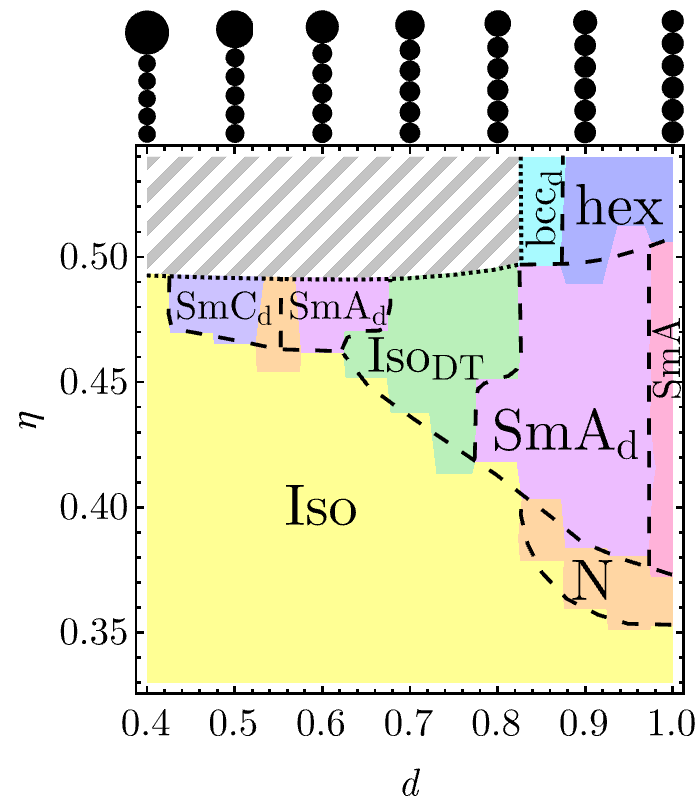}
    \caption{Phase diagram in the phase space of smaller balls' diameter $d$ and packing fraction $\eta$. The following phases are present: isotropic (Iso), uniform nematic (N), type A smectic (SmA), interdigitated type A smectic ($\text{SmA}_\text{d}$), interdigitated type C smectic ($\text{SmC}_\text{d}$), isotropic double-twisted cluster liquid ($\text{Iso}_\text{DT}$), interdigitated bcc crystal ($\text{bcc}_\text{d}$) and non-polar hexatic crystal (hex). The hatched filling marks the region that was not reached in the simulations due to computational limits.}
    \label{fig:pd}
\end{figure}

\begin{figure}[htb]
    \centering
    \includegraphics[width=0.48\linewidth]{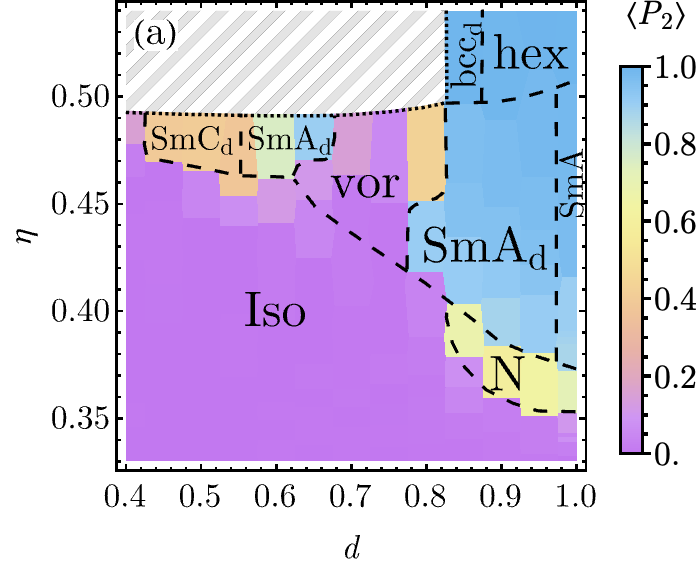}
    \includegraphics[width=0.48\linewidth]{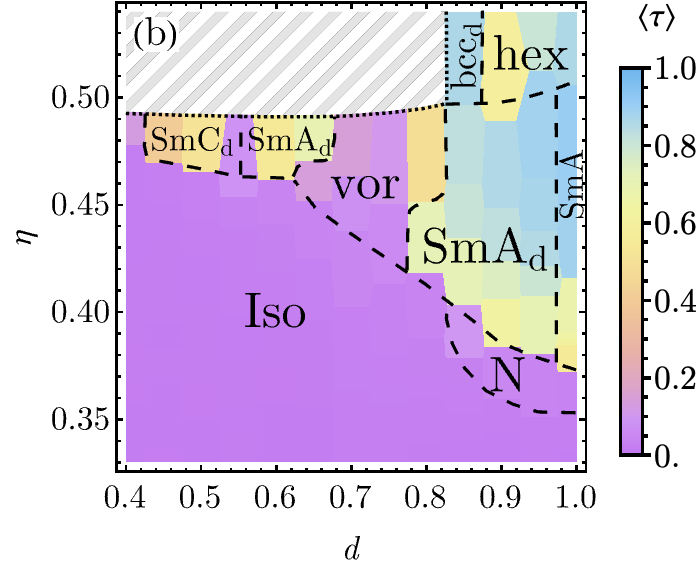}
    \includegraphics[width=0.48\linewidth]{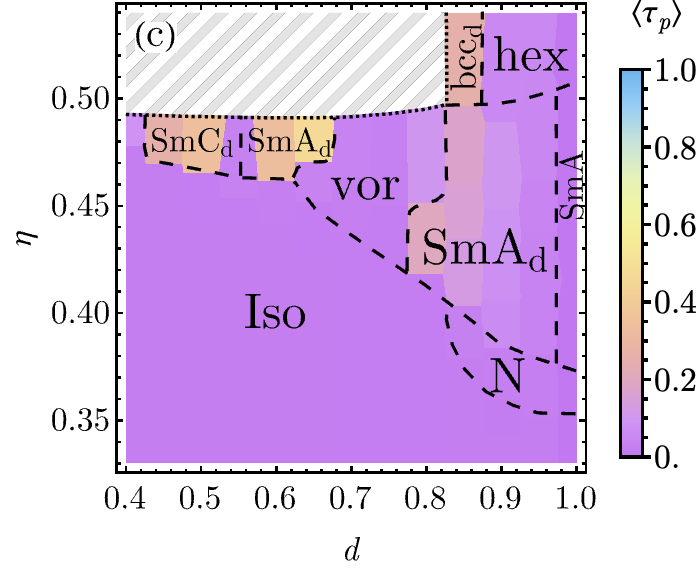}
    \includegraphics[width=0.48\linewidth]{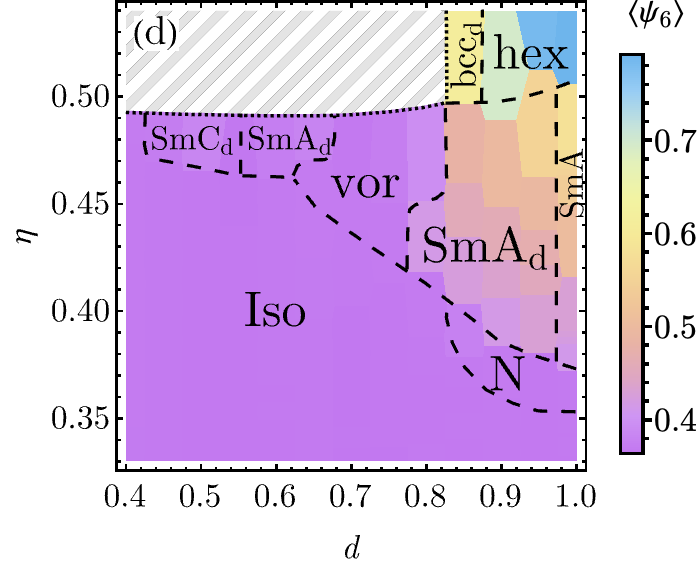}
    \caption{The dependence of selected observables on the smaller balls' diameter $d$ and packing fraction $\eta$: (a) nematic order parameter $\expval{P_2}$, (b) smectic order parameter $\expval{\tau}$, (c) polarization order parameter $\expval{\tau_p}$, and (d) local hexatic bond order parameter $\expval{\psi_6}$. The hatched filling marks the region that was not reached in the simulations due to computational limits.}
    \label{fig:obs}
\end{figure}

\begin{figure}[htbp]
    \centering
    \includegraphics[width=0.8\linewidth]{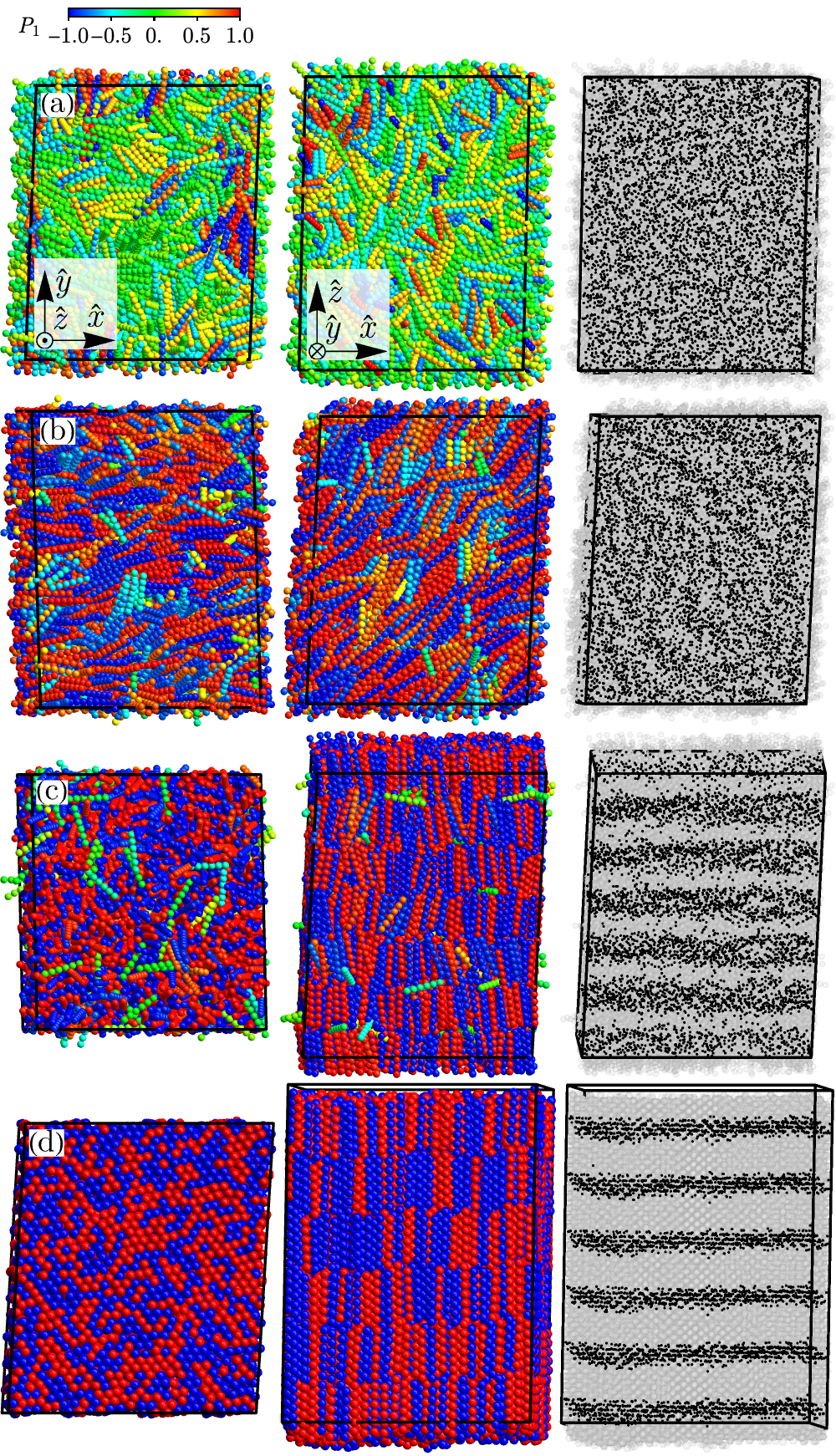}
    \caption{System snapshots of the phase sequence in a system of linear tangent hard sphere (LTHS) hexamers ($d = 1$), common for elongated particles. Left and middle columns are, respectively, top and side views of the simulation box; the bottom edge of the box in the top view (left column) is the top edge in the side view (middle column). The right column shows the side view with geometric centers (black dots). The particles are color-coded according to $P_1 = \vu{n} \vdot \vu{a}$. Subsequent rows show the compression sequence: (a) Iso ($\eta = 0.34$), (b) N ($\eta = 0.36$), (c) SmA ($\eta = 0.39$), and (d) hex ($\eta = 0.54$).}
    \label{fig:pack_1p0}
\end{figure}

\begin{figure}[htbp]
    \centering
    \includegraphics[width=0.75\linewidth]{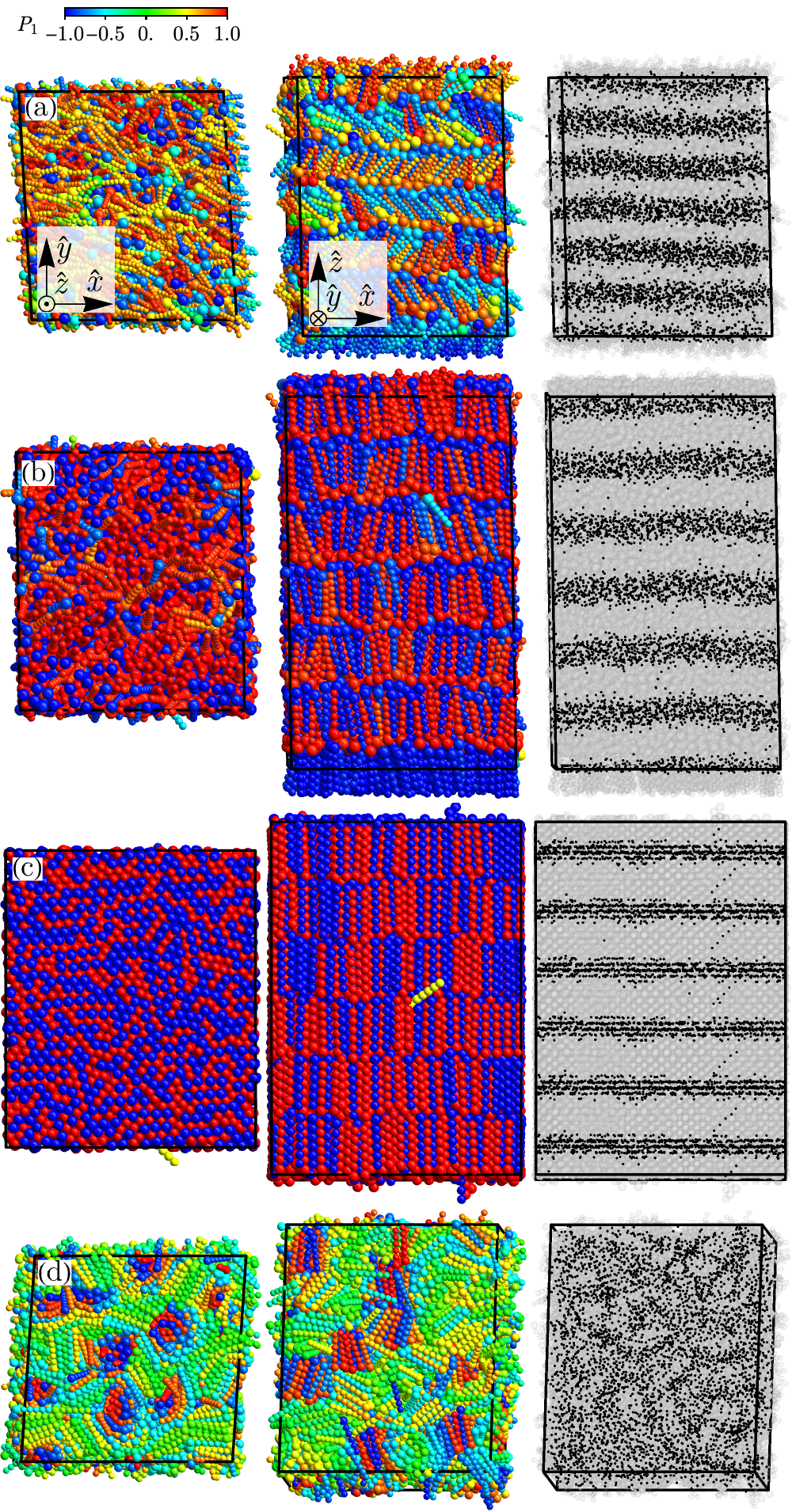}
    \caption{Snapshots of phases different than the ones observed for $d=1$ (Fig.~\ref{fig:pack_1p0}). The panel follows the same layout as Fig.~\ref{fig:pack_1p0} (columns, from the left: top view, side view, side view of geometric centers). Subsequent rows correspond to: (a) $\text{SmA}_\text{d}$ [$(d, \eta) = (0.65, 0.48)$], (b) $\text{SmC}_\text{d}$ [$(d, \eta) = (0.5, 0.48)$], (c) $\text{Iso}_\text{DT}$ [$(d, \eta) = (0.75, 0.46)$], and (d) $\text{bcc}_\text{d}$ [$(d, \eta) = (0.85, 0.52)$].}
    \label{fig:pack_nice}
\end{figure}

Using the observables and correlation functions described in Section~\ref{sec:params}, as well as the visual inspection of final system snapshots, several phases were recognized in the system. These include: isotropic liquid (Iso), uniform nematic (N), type A smectic (SmA), interdigitated type A smectic ($\text{SmA}_\text{d}$), interdigitated type C smectic ($\text{SmC}_\text{d}$), isotropic double-twisted cluster fluid ($\text{Iso}_\text{DT}$), interdigitated bcc crystal ($\text{bcc}_\text{d}$) as well as hexatic crystal (hex). Types of phase transitions (e.g., first order, second order) were not recognized due to long simulation times. The phase diagram is shown in Fig.~\ref{fig:pd}, while selected order parameters as a function of system parameters are presented in Fig.~\ref{fig:obs}. Finally, Figs.~\ref{fig:pack_1p0} and \ref{fig:pack_nice} show final system snapshots of all recognized phases; among them, Fig.~\ref{fig:pack_1p0} depicts the standard phase sequence observed for elongated particles, while Fig.~\ref{fig:pack_nice} gathers less common ones.

\subsection{The standard phase sequence}

Typically, for elongated particles, one observes the Iso-N-SmA-crystal phase sequence on compression route. Here, it is also the case for $d = 1$ (see Fig.~\ref{fig:pd}) constituting for linear tangent hard sphere (LTHS) hexamer \cite{Vega1992, Vega2001, Kubala2023}, where the radii are identical. The following paragraphs briefly summarize the extent and characteristics of these standard phases.

The isotropic liquid (Fig.~\ref{fig:pack_1p0}a) is observed in a wide range of system parameters. Its border falls monotonically with growing $d$. For $d=0.4$, no other phase was observed for packing fractions less than $\eta = 0.5$. For $d=1.0$, Iso phase is stable until $\eta = 0.35$, where the Iso-N phase transition occurs. All order parameters are close to their disorder value (see Fig.~\ref{fig:obs}). One can observe a slight increase of nematic $\expval{P_2}$ and smectic $\expval{\tau}$ order parameters (Fig.~\ref{fig:obs}a,b) near the phase boundary.

The nematic phase (Fig.~\ref{fig:pack_1p0}b), adjacent to Iso, occurs only in a narrow range of $d$ and $\eta$. For $d \le 0.85$ it is completely absent. It is signified by a large value of the nematic order parameter $\expval{P_2} \approx 0.6-0.7$, being near the upper limit of a typical range $\expval{P_2} \in [0.3, 0.7]$ observed in the experiments. Comparable values of $\expval{P_2}$ were measured in analogous studies of tapers \cite{Kubala2023, Kubala2023sm}. The smectic order is near zero. A quick destabilization of the N phase with decreasing $d$ is surprising, as for $d=0.8$ the particle is visually almost indistinguishable for the $d=1$ case (\textit{cf}. Fig.~\ref{fig:pd}). On one hand, LTHS hexamer is the first of the LTHS $k$-mer family which stabilizes the nematic phase (for $k \le 5$ SmA is formed directly over Iso), thus the N phase may be susceptible to slight variations of particle's shape. On the other hand, in the study of tapers \cite{Kubala2023}, the N phase appears even for shapes with a high anisotropy of balls' radii.

A standard type A smectic appears over the N phase only for $d = 1$ (Fig.~\ref{fig:pack_1p0}c); for lower $d$ values, all smectics exhibit a varying but non-zero degree of interdigitation, which is discussed later. Nematic order is nearly perfect with $\expval{P_2}$ close to 1. Smectic order is significant, which is confirmed by $\expval{\tau}$ increasing from $\expval{\tau} \approx 0.55$ to nearly 1 upon compression. One also observes a rise of local hexatic order signified by $\expval{\psi_6}$ reaching $\approx 0.58$ near the upper phase boundary; the preference of having six neighbors facilitates more optimal local packing, however inspection of system snapshots does not reveal long-range hexatic order, as, e.g., in the SmB phase \cite{Birgeneau1978}. Polarization order $\expval{\tau_p}$ is, by definition, zero, since the particle is non-polar.

A non-polar hexatic crystalline phase is observed over $\eta = 0.5$ for $d \ge 0.9$ (Fig.~\ref{fig:pack_1p0}d). High local hexatic order is confirmed by values of $\expval{\psi_6} > 0.7$ and reaching close to 1 for $d = 1$. Moreover, as it can be inferred from system snapshots, the hexatic order is actually long-range. The hex phase is formed by stacking hexagonal honeycomb layers on top of one another, with at most local correlations of particles' polarization. The values of $\expval{P_2}$ and $\expval{\tau}$ are high, although $\expval{\tau}$ falls to $\approx 0.6$ for $d = 0.9$. In this case, the system slightly resembles the hexatic columnar $\text{Col}_\text{h}$ phase \cite{Grelet2008}. However, the columnar phase remains a liquid in the direction of columns, and $\expval{\tau}$ consequently falls to 0, while here the system is jammed, and a non-zero equilibrium value of $\expval{\tau}$ is reached.

\subsection{Interdigitated phases} \label{sec:interdigitated}

\begin{figure}[htb]
    \centering
    \includegraphics[width=0.5\linewidth]{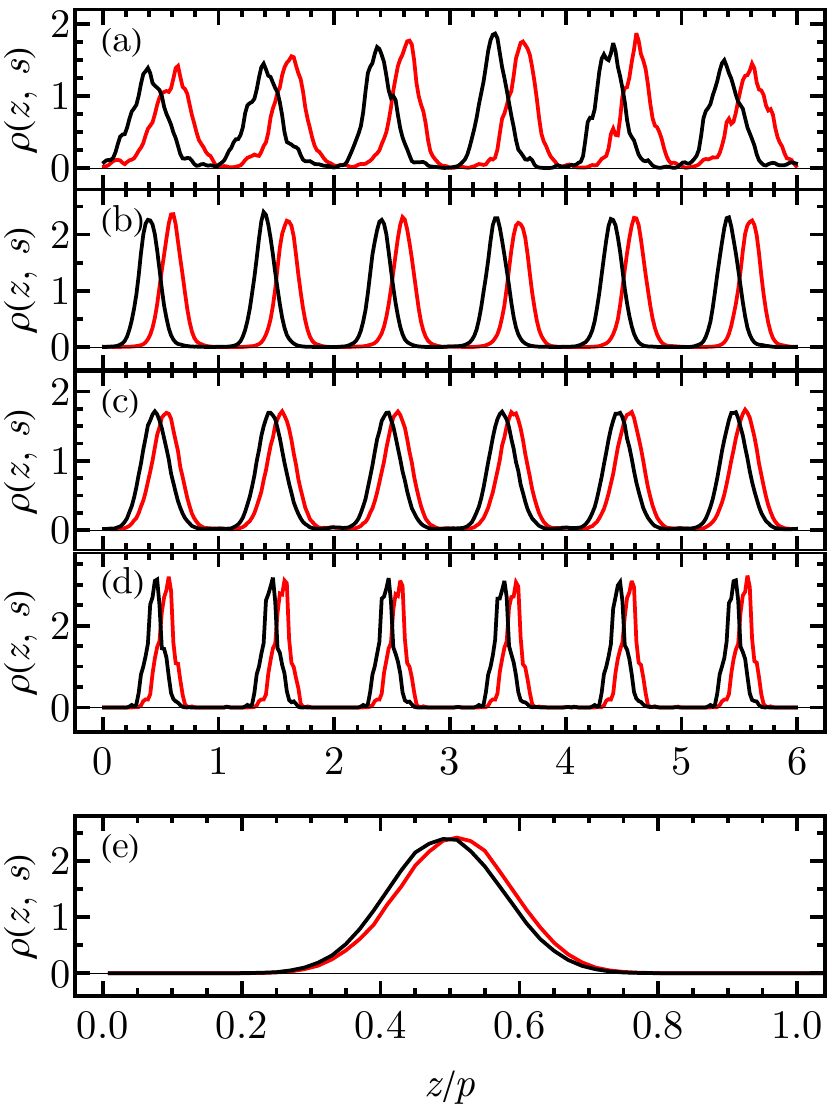}
    \caption{Normalized particle density $\rho(z, +1)$ of ``up'' particles (red line) and $\rho(z, -1)$ of ``down'' particles (black line) as a function of $z$ coordinate rescaled by the layers' width $p$ (the period) for (a) $\text{SmC}_\text{d}$ [$(d,\eta)=(0.55,0.48)$], (b) $\text{SmA}_\text{d}$ [$(d,\eta)=(0.65,0.48)$], (c) $\text{SmA}_\text{d}$ [$(d,\eta)=(0.8,0.44)$], (d) $\text{bcc}_\text{d}$ [$(d,\eta)=(0.85,0.52)$], and (e) $\text{SmA}_\text{d}$ [$(d,\eta)=(0.95,0.45)$]. In the last (e) panel, only a single period is shown}
    \label{fig:rho}
\end{figure}

\begin{figure}[htb]
    \centering
    \includegraphics[width=0.6\linewidth]{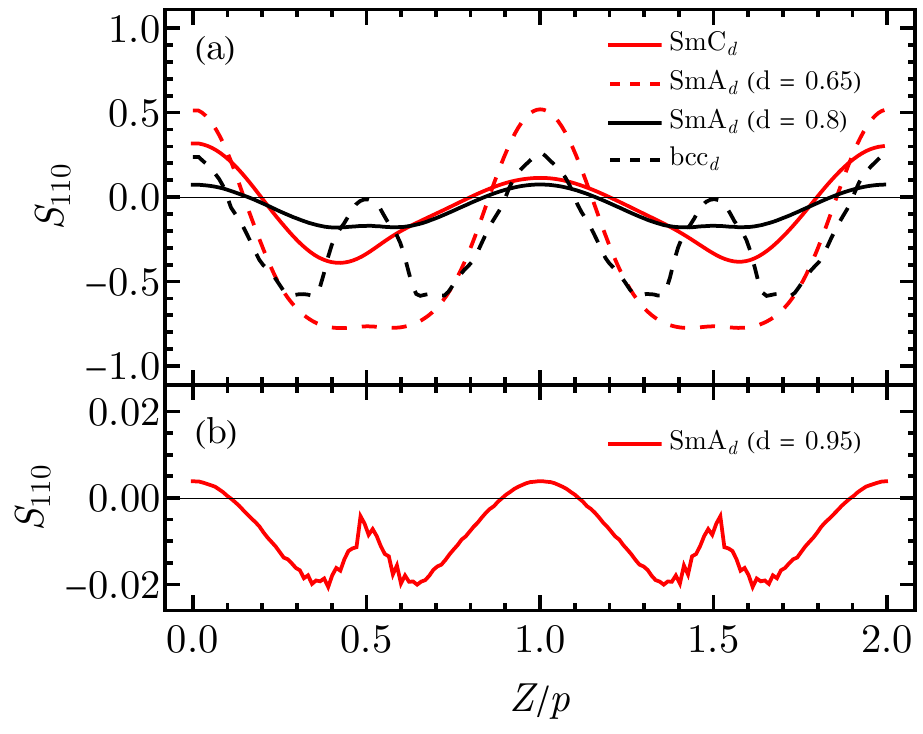}
    \caption{The dependence of $S_{110}$ correlation function on the distance $Z$ along the $z$ axis rescaled by the period $p$ (layers' width) for different system parameters. Only the first two $Z \in [0, 2p]$ of the six periods fitting in the box are shown; two further period pairs $Z \in [2p, 4p]$ and $Z \in [4p, 6p]$ are nearly identical. (a) The dependence for: red solid line -- $\text{SmC}_\text{d}$ [$(d,\eta)=(0.55,0.48)$], red dashed line -- $\text{SmA}_\text{d}$ [$(d,\eta)=(0.65,0.48)$], black solid line -- $\text{SmA}_\text{d}$ [$(d,\eta)=(0.8,0.44)$], black dashed line -- $\text{bcc}_\text{d}$ [$(d,\eta)=(0.85,0.52)$]. (b) The dependence for $\text{SmA}_\text{d}$ [$(d,\eta)=(0.95,0.45)$] on a magnified scale. Here, an additional averaging is performed over three period pairs $Z \in [Kp, (K+1)p]$, $K=0,2,4$, for improved statistics.}
    \label{fig:s110}
\end{figure}

For $d < 1$, the particle ceases to possess the up-down symmetry. Consequently, polar phases may emerge. Phases with a bulk, non-zero polarization were recently discovered for the RM734 molecule \cite{Sebastian2020}, among which the most important one is the ferroelectric nematic $\text{N}_\text{F}$. It is widely believed that the most important molecular factors responsible for stabilizing the $\text{N}_\text{F}$ phase are the taper-like shape of the mesogen as well as a large dipolar moment of the molecule \cite{Mandle2021}. Another polar phase appearing in the phase sequence of RM734 is splay nematic $\text{N}_\text{S}$, where the bulk polarization is zero, however, one observes alternating areas with non-zero polarization and a presence of splay modulation. De Gregorio \textit{et al.} \cite{Gregorio2016} argued, basing on the second-order density functional theory (DTF), that a taper-like shape of the particle is enough to promote long-range splay in the system. The shape effect has been studied numerically by our research group in two recent papers \cite{Kubala2023, Kubala2023sm} treading similar particles as in the current study built of six or eleven hard tangent balls, however, with linearly increasing radii (notably, the eleven-ball model was identical as in the DFT study \cite{Gregorio2016}). It was shown that for such particles, entropic interactions are not enough to stabilize either $\text{N}_\text{F}$ or $\text{N}_\text{S}$ phases.

Similarly, in our system of lollipops, neither nematic nor smectic ferroelectric or splay phases are observed. However, antiferroelectric, interdigitated smectics $\text{SmA}_\text{d}$ and $\text{SmC}_\text{d}$, as well as $\text{bcc}_\text{d}$ crystal appear. A polar interdigitated smectic is formed by superimposing two phase-shifted smectics with the same mean density and period but with different polarizations \cite{Blatch1995}. In first order, the density profile $\rho(z, s)$ can be described as 
\begin{align}
    \rho(z, +1) &= \frac{\rho}{2} + \frac{\rho}{2} \tilde{\tau} \sin(k z), \\
    \rho(z, -1) &= \frac{\rho}{2} + \frac{\rho}{2} \tilde{\tau} \sin[k (z - \Delta z)], \\
\end{align}
where the ideal nematic order is assumed, $s = \vu{a}_i \vdot \vu{n} = \pm 1$, $\rho = \eta/V_P$ is the mean number density, $\tilde{\tau} \in [0, 1]$ controls the depth of undulations, and $\Delta z$ is the distance between the maximal densities for opposite polarization. The total density profile is then equal
\begin{equation}
    \rho(z) = \sum_{s = \pm 1} \rho(z,s) = \rho + \rho \tau \sin\qty[k\qty(z - \frac{\Delta z}{2})],
\end{equation}
where $\tau = \tilde{\tau} \cos(k \Delta z/2)$. The structure has a non-zero value of the polarization order parameter
\begin{equation}
    \tau_p = \frac{1}{\rho p} \abs{\sum_{s = \pm 1} \int_{0}^{p} \dd{z} s \exp(\iota k z) \rho(z,s)} = \frac{\tilde{\tau}}{2} \abs{\sin(\frac{k\Delta z}{2})},
\end{equation}
where $p = 2\pi/k$ is the layer width (the period). Clearly, the value of $\tau_p$ grows with $\Delta z$ and is zero for $\Delta z = 0$, when the interdigitation disappears.

One of the interdigitated phases, the type C interdigitated smectic $\text{SmC}_\text{d}$ appears in the range $d \in [0.45, 0.5]$ for packing fractions $\eta > 0.47$. Its snapshot is presented in Fig.~\ref{fig:pack_nice}a. The director $\vu{n}$ within each layer is tilted from the smectic wavevector $\vb{k}$, $\vu{n} \nparallel \vb{k}$. The tilt angle is nearly constant, but the direction alternates in the neighboring layer, which means that the smectic is anticlinic ($\text{SmC}_\text{A})$ \cite{Chandani1989, Osipov2000}. It is reflected in a significantly smaller value of the nematic order parameter $\expval{P_2} \approx 0.4$ (Fig.~\ref{fig:obs}a) than in other smectic and crystalline phases. Each layer separates into two components with opposite polarizations. The interdigitation can be clearly seen in the snapshot as well as in Fig.~\ref{fig:rho}a, where the numerical, ensemble-averaged density profiles $\rho(z,s)$ are presented separately for $s=+1$ and $-1$. The separation $\Delta z$ is large, near one-third of the period $p$. The density peaks are not identical for all layers. Most probably, it can be attributed to sporadic structure defects visible in some of the layers (\textit{cf}. Fig.~\ref{fig:pack_nice}a) and fluctuations of the tilt angle. Interdigitation is further signified by a non-zero value of the polarization order parameter $\expval{\tau_p} \approx 0.35$ (Fig.~\ref{fig:pack_nice}c). Last but not least, we observe a non-trivial profile of the $S_{110}$ correlation function (red solid line in Fig.~\ref{fig:s110}a). The maxima for the distances $Z = 2Kp$, $K \in \mathbb{Z}$ are equal $\approx 0.3$ and are higher than for $Z = (2K+1)p$, which are equal $\approx 0.15$; the first set corresponds to correlations between the layers with an identical tilt, while the second one -- to opposite tilts, which explains the lower value. In between the maxima, there are minima with $S_{110} \approx -0.35$ for $Z \approx (2K + 0.4)p, (2K + 1.6)p$, which further proves the existence of sub-layers with a different polarization.

Type A interdigitated smectic $\text{SmA}_\text{d}$ (see Fig.~\ref{fig:pack_nice}b for a system snapshot) is observed in two regions of the phase diagram, separated by another phase (isotropic double-twisted cluster liquid, discussed in the next subsection). The smaller region appears for $d \in [0.6, 0.65]$ and $\eta > 0.46$, while the second, larger one for $d \in [0.8, 0.95]$ and $\eta > 0.38$. Here, the director $\vu{n}$ is parallel to the smectic wavevector $\vb{k}$, $\vu{n} \parallel \vb{k}$ and both nematic order $\expval{P_2}$ and smectic order $\expval{\tau}$ parameters have relatively large values ($\expval{P_2} > 0.75$ and $\expval{\tau} > 0.55$). Alike the $\text{SmC}_\text{d}$ phase, the interdigitation is reflected in the observables and the correlation functions. Fig.~\ref{fig:rho}b,c,e presents the density distribution $\rho(z, s)$ for one point in the $\text{SmA}_\text{d}$ region for lower $d$ values and two points in the second $\text{SmA}_\text{d}$ region for higher $d$ values. In all three cases, the density peaks are smooth and nearly identical for all six layers. The separation $\Delta z$ falls with increasing $d$, with $\Delta z \approx 0.2p$ for $d = 0.65p$, $\Delta z \approx 0.15$ for $d = 0.8$ and $\Delta z \approx 0.03p$ for $d = 0.95$. It is worth noting that even for the highest $d < 1$ studied, $d = 0.95$, the interdigitation minuscule as it may be, it is clearly visible over the statistical fluctuations (which are smaller than the line width in Fig.~\ref{fig:rho}e). It leads us to believe that the $\text{SmA}_\text{d}$ phase may be present for each $d < 1$, regardless of how close to 1 the $d$ is. The polarization order parameter (Fig.~\ref{fig:obs}c) is visibly larger than 0 for all $d < 0.95$ and mostly rises with falling $d$ (excluding the $d = 0.6$ case), reaching $\expval{\tau_p} \approx 0.5$ for $d = 0.65$. For $d=0.95$, its value is indistinguishable from the non-polar case within statistical error bounds. The analysis of $S_{110}$ correlation function for the same points (red dashed and black solid lines in Fig.~\ref{fig:s110}a, as well as the only line in Fig.~\ref{fig:s110}b) also confirms the interdigitation pattern. All three cases presented are qualitatively similar -- the maxima appear for $\Delta Z = Kp$, $K \in \mathbb{Z}$ and deeper, wider valleys originating from the opposite polarizations are placed around $\Delta Z = (K+0.5)p$. Exactly in $\Delta Z = (K+0.5)p$, one can observe additional, small maxima. As it can be seen, e.g., in the snapshot from Fig.~\ref{fig:pack_nice}b, there are hardly any particles between the layer, and if they appear, they are usually tilted w.r.t. the director. In such a case, $S_{110}$ is higher than for the antiparallel alignment, explaining the presence of these small maxima. They are visible most clearly for $d=0.95$ (Fig.~\ref{fig:s110}b), where the overall values of $S_{110}$ are close to 0; one can also see significantly larger fluctuation in these regions in comparison to $\Delta Z \approx Kp$. Lastly, similarly to the standard SmA smectic for $d = 1$, for $d > 0.8$, a rise of the local hexatic order parameter $\expval{\psi_6}$ is observed (Fig.~\ref{fig:obs}d). However, once again, hexatic correlations are at most local.

Between the $\text{SmA}_\text{d}$ and $\text{SmC}_\text{d}$ phases, for $d = 0.55$, the system fails to create a smectic phase, and instead a high-density nematic is formed (as signified by a high $\expval{P_2}$ and a low $\expval{\tau}$ values, c.f. Fig.~\ref{fig:obs}a,b). The layering present in the initial configuration is destroyed during the course of a simulation. Inspection of the simulation trajectory (results not shown) reveals that small, local smectic clusters are formed. This hints that the culprit may be the competition between type A and C smectics, in which the angles between the director $\vu{n}$ and the wavevector $\vb{k}$ are different. However, one can never exclude the possibility that the system may have become jammed in a metastable configuration, requiring significantly more simulation cycles to relax and rebuild well-defined layers, which is beyond our computational limits.

Finally, we present the reasoning why one observes the SmC phase for smaller $d$ values and SmA for larger ones, as well as what is the origin of the interdigitation. Both effects may be attributed to the packing efficiency. The mean distance between the particles is imposed by the largest balls. Then, however, the chains of smaller ones are packed suboptimally. The remedy is the superposition of two oppositely polarized structures. That way, there is two times more chains to pack in the layer. Moreover, all large balls are in contact, which helps them to arrange more optimally among themselves. For small $d$ values, this mechanism is not sufficient enough. However, introducing the tilt mitigates the issue -- the layers' width and, consequently, their volume is decreased, allowing once again for a better packing of the chains. The same effect was observed in the two-dimensional system of lollipops \cite{Perez2017}.

The last phase with the interdigitation of oppositely polarized substructures is the tetratic crystalline phase (Fig.~\ref{fig:pack_nice}c). It is formed by superimposing two tetratic lattices with opposite polarizations, shifted in $X$ and $Y$ direction by half of the unit cell size. Thus, we call the structure \textit{interdigitated body-centered cubic crystal} ($\text{bcc}_\text{d}$), although one has to remember that the distances between interdigitated counterparts are asymmetric in the $z$ direction. Interdigitation is confirmed by the $\rho(z, s)$ density profile (Fig.~\ref{fig:rho}d), non-zero value of $\expval{\tau_p}$ as well as $S_{110}$ correlation function (black dashed line in Fig.~\ref{fig:s110}a). In the case of the $S_{110}$ function, well pronounced maxima for $\Delta Z = (K+0.5)p$ can be probably again attributed to the outliers with anomalous orientations placed in between the layers. Tetratic order is confirmed by the local tetratic order parameter $\expval{\psi_4} \approx 0.72$ (result not shown), which is significantly larger than for all other simulation points. Hexatic order parameter also has a significant value $\expval{\psi_6} \approx 0.65$. This phenomenon can be explained by noticing that the projection of bcc structure on a plane essentially creates flattened hexagons, which is consequently detected by $\expval{\psi_6}$.

\subsection{The isotropic double-twisted cluster liquid}

\begin{figure}[htb]
    \centering
    \includegraphics[height=0.35\linewidth]{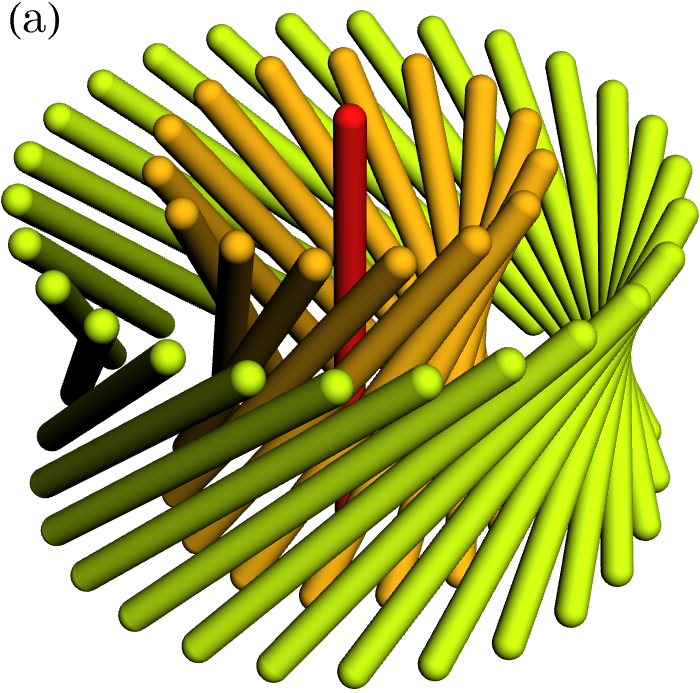}
    \includegraphics[height=0.35\linewidth]{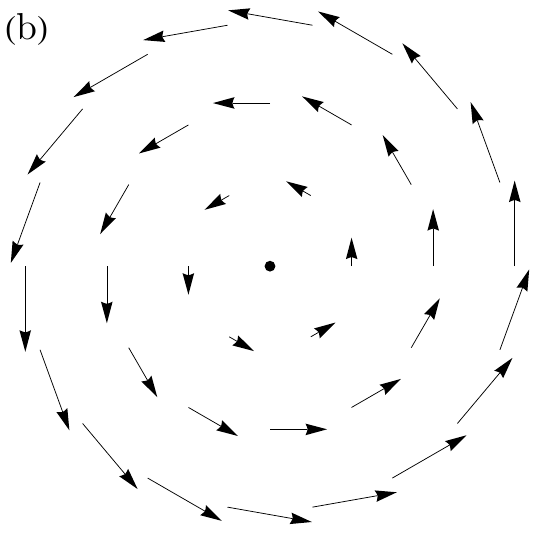}
    \caption{(a) Illustrative representation of the preferred directions in the cluster. The tilt of the director increases with the distance from the middle. (b) Projections of the director on a cluster plane. The field lines of an ideal field form concentric circles around the cluster center $\vb{r}_c$. In such a configuration, the curl order parameter $\chi$ [Eq.~\eqref{eq:chi}] reaches its maximal magnitude $\abs{\chi} = 1$.}
    \label{fig:vortex}
\end{figure}

\begin{figure}[htb]
    \centering
    \includegraphics[height=0.35\linewidth]{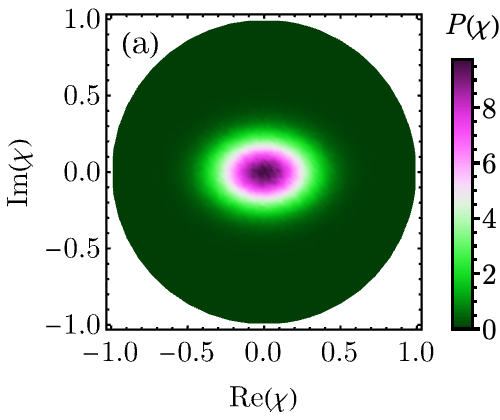}
    \includegraphics[height=0.35\linewidth]{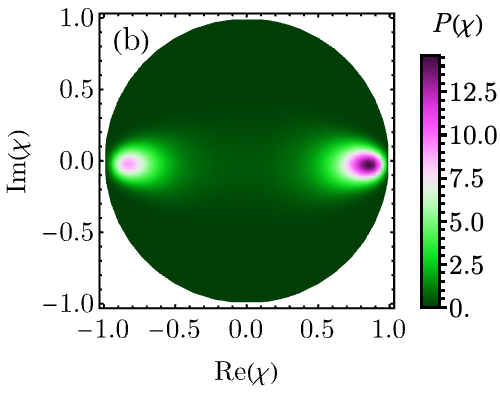}
    \caption{Histograms of the curl order parameter $\chi$ on a complex plane for selected parameter sets: (a) Iso [$(d, \eta) = (0.75, 0.35)$] and (b) $\text{Iso}_\text{DT}$ [$(d, \eta) = (0.75, 0.46)$].}
    \label{fig:chi}
\end{figure}

\begin{figure}[htb]
    \centering
    \includegraphics[width=0.55\linewidth]{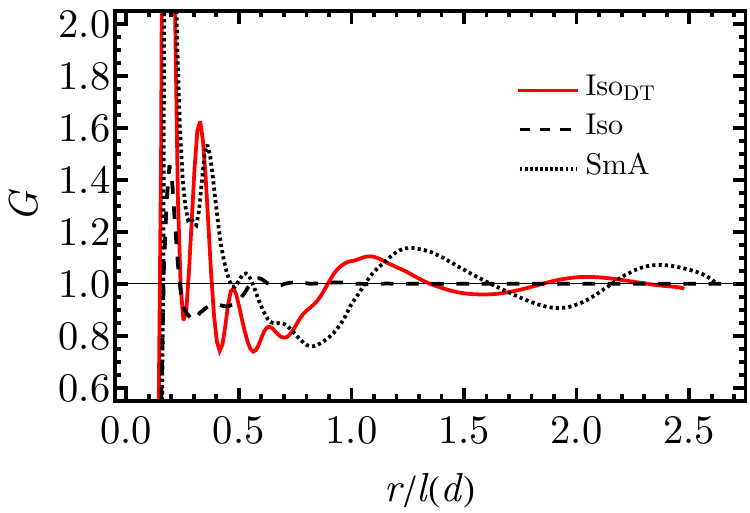}
    \caption{The dependence of the radial distribution function $G(r)$ on the distance $r$ in the units of the particles length $l(d)$ for selected phases: red solid line -- $\text{Iso}_\text{DT}$ [$(d, \eta) = (0.75, 0.46)$], black dashed line -- Iso [$(d, \eta) = (0.75, 0.35)$], and black dotted line -- SmA [$(d, \eta) = (1, 0.39)$]. Horizontal black line marks the value $G(\infty) = 1$.}
    \label{fig:G}
\end{figure}

The two $\text{SmA}_\text{d}$ regions are separated by a phase for which all ensemble-averaged values of order parameters ($\expval{P_2}$, $\expval{\tau}$, $\expval{\tau_p}$ and $\expval{\psi_6}$) in Fig.~\ref{fig:obs} are close to their disorder values. This suggests that the phase is globally isotropic. However, the inspection of the system snapshot (Fig.~\ref{fig:pack_nice}d) reveals some mesostructure. The system consists of small clusters with a size comparable to the particle's length. Each cluster has a local, preferred direction $\vu{n}_c$. We will be calling the plane perpendicular to $\vu{n}_c$ the \textit{cluster plane}. The analysis of clusters' structure reveals that the long molecular axes $\vu{a}_i$ are tilted in the direction tangent to a family of circles lying in the cluster plane, centered around the cluster center $\vb{r}_c$. Moreover, the tilt increases with the distance from $\vb{r}_c$. The sense of the axes $\vu{a}_i$ is random, thus, the phase appears to be apolar. The structure of the director field in the cluster is illustrated in Fig.~\ref{fig:vortex}a. Consequently, the projections $\tilde{\vb{a}}_i = \vu{a}_i - \vu{n}_c (\vu{n}_c \vdot \vu{a}_i)$ of the axes $\vu{a}_i$ of the particles surrounding the center $\vb{r}_c$ are tangent to the same family of circles and the length of $\tilde{\vb{a}}_i$ vectors increases with the distance from $\vb{r}_c$. The projections are illustrated in Fig.~\ref{fig:vortex}b.

The director field modulation described above is known in the literature as \textit{double-twist} \cite{Nayani2015,Revignas2023}. Double-twist is observed in many systems, among which the most known are the blue phases (BP). The director field in type I, II and III BP forms long, double twisted tubes with linear topological defects between them (the \textit{disclinations}). While disclinations in type I and II BP form a regular lattice, type III is disordered, which bears the most resemblance to our system. The main difference is the length scale -- while the size of our clusters is comparable to the particles' length, the modulations in blue phases are usually on the scale of hundreds of particles.

To reflect the most important features of the phase described in this section, we call it the \textit{isotropic double-twist cluster liquid} ($\text{Iso}_\text{DT}$). The double-twisted clusters appearing in $\text{Iso}_\text{DT}$ are readily detected by the curl order parameter $\chi$ (\ref{eq:chi}). Its histogram on a complex plane is shown in Fig.~\ref{fig:chi}. Panel \ref{fig:chi}a shows $\chi$ for the standard isotropic liquid. As expected, $\chi$ is close to 0 for most particles. The spread of the peak can be attributed to statistical fluctuation, due to which a local double-twisted structure may be momentarily formed. The situation changes in the $\text{Iso}_\text{DT}$ phase (Fig.~\ref{fig:chi}b). Here, the system's most probable values of $\chi$ 
are $\approx \pm 0.8$. It means that most particles belong to double-twisted clusters. As the preferred values of $\chi$ lie on the real axis, it is confirmed that the tilt direction (the $\tilde{\vb{a}}_i$ projected vector) is tangent to the concentric circles. The values of $\chi$ with an opposite complex phase correspond to clusters with a different chirality (relative to the central particle's polarization).

A closer inspection of the $\text{Iso}_\text{DT}$ phase system snapshot (Fig.~\ref{fig:pack_nice}d) reveals a local density modulation. The density profile is, however, not periodic as for smectics observed in the system in other regions of the phase diagram. To gain more insight into the density modulation, we computed the radial distribution function $G(r)$ \cite{Stone1978}. The comparison of $G(r)$ for $\text{Iso}_\text{DT}$, Iso and SmA phases is presented in Fig.~\ref{fig:G}. For the Iso phase (black dashed line), we observe a few short-ranged maxima and minima separated by $\Delta r \approx 0.2 l(d)$, where $l(d)$ is the length of the particle, which roughly corresponds to the width of the particle. The modulations vanish before $r$ reaches $l(d)$ value. On the other hand, the SmA phase (black dotted line) behaves differently. Short-ranged minima and maxima with $\Delta r \approx 0.2 l(d)$ separation are significantly more pronounced. Moreover, a second series of maxima appears with separation $\Delta r \approx 1.2 l(d)$, which originates from the layering of the system. By comparing the $G(r)$ profile of $\text{Iso}_\text{DT}$ (red solid line) with the two other phases, one clearly sees that it is closer to the SmA case. In particular, we observe secondary maxima with the separation $\Delta r \approx l(d)$. This means that the  $\text{Iso}_\text{DT}$ phase may be described as locally smectic-like.

The existence of the $\text{Iso}_\text{DT}$ phase can be explained by the observation that in the nematic and smectic phases, the effective shape of the elongated particles thermally vibrating around the director resembles an hourglass \cite{Revignas2023}. The double-twisted configuration is a natural way to pack particles with such a shape. In the system of lollipops, the hourglass shape effect may be amplified by the formation of pairs of lollipops with an opposite polarization. Moreover, the tilt inherent to the double-twist, as we discussed in Sec.~\ref{sec:interdigitated}, increases the packing efficiency of small molecules' chains. The $\text{Iso}_\text{DT}$ phase is in direct competition with the type A interdigitated smectic $\text{SmA}_\text{d}$. Both phases rely on different mechanisms to increase the packing entropy. The ones providing a larger gain decide which of the two phases is eventually formed.

\section{Summary and outlook}

In this manuscript, we have analyzed the system of hard particles built of six tangent balls, whose shape resembles a lollipop. The particles modeled the RM734 molecule, for which ferroelectric and splay polar nematics were observed in recent experiments. The systems were simulated using the isothermal-isobaric Monte Carlo sampling in the orthorhombic as well as the general triclinic simulation boxes. We varied two independent parameters: the packing fraction $\eta$ and the diameter of the smaller balls $d$, while keeping the diameter of the largest ball constant and equal one. The phases present in the system were identified by inspection of system snapshots, together with the computation of expectation values of order parameters, such as the nematic order parameter $\expval{P_2}$, smectic order parameter $\expval{\tau}$, polarization order parameter $\expval{\tau_p}$, bond order parameters $\expval{\psi_4}$, $\expval{\psi_6}$, and curl order parameter $\expval{\tau}$, as well as $S_{110}$ correlation function, radial distribution function $G(r)$, the density profile $\rho(z, s)$, and histograms of the curl order parameter $\chi$. Apart from a standard phase sequence Iso-N-SmA-crystal, commonly observed for elongated particles, we identified four other phases. The first group was three interdigitated phases: type A and C interdigitated smectics ($\text{SmA}_\text{d}$ and $\text{SmC}_\text{d}$, respectively) as well as interdigitated body centered cubic crystal ($\text{bcc}_\text{d}$). All of them were composed of two sets of oppositely polarized smectic or crystalline layers with a slight spatial dephasing between them, which in consequence made those phases polar. It is worth noting that for a similar system with taper-like particles \cite{Kubala2023,Kubala2023sm}, no liquid crystalline polar phases were observed. The fourth, and arguably the most intriguing phase, was the isotropic double-twist cluster liquid ($\text{Iso}_\text{DT}$). It was built of small clusters of particles twisting around the center; the clusters were oriented isotropically and spatially uncorrelated, resembling type III blue phase, but on a much smaller scale. The phase was locally smectic-like, however, the density modulations did not create any periodic pattern.

The manuscript contributed to a better understanding of the shape effect of particles with a broken up-down symmetry, which has been of particular interest in recent years. This model has a potential for further expansions, which include identifying the phases in the regions of the phase diagram not reached in this study, as well as exploration of its other variants, such as a different number of spheres or a smooth surface. Moreover, based on the observation that the phase diagrams of lollipops and tapers are vastly different, studying the self-assembly of other particles with a broken up-down symmetry may result in the observation of many novel phases.

\section*{Data availability}

The datasets generated during and/or analyzed during the current study are available from P.K. upon reasonable request.

\section*{Code availability} \label{sec:code}

The source code of an original RAMPACK simulation package used to perform Monte Carlo sampling is available at \url{https://github.com/PKua007/rampack}.

\section*{Acknowledgements}
Through this paper, which appears as part of a special issue dedicated to Prof. Lech Longa, we express our deep gratitude for showing us the fascinating world of liquid crystals, guidance, mentorship, and many inspiring discussions and ideas he shared with us.

P.K. acknowledges the support of Ministry of Science and Higher Education (Poland) grant no. 0108/DIA/2020/49. M.C. acknowledges the support of National Science Center in Poland grant no. 2021/43/B/ST3/03135. Numerical simulations were carried out with the support of the Interdisciplinary Center for Mathematical and Computational Modeling (ICM) at the University of Warsaw under grant no. G27-8.

\section*{Funding}

P.K. acknowledges the support of the Ministry of Education and Science (Poland) grant no. 0108/DIA/2020/49.

\bibliographystyle{tfnlm} 
\bibliography{ref}

\end{document}